\documentstyle[iopconf1,epsf]{article}
\begin{document}
\title{Big Bang Nucleosynthesis: Reprise}
\author{Subir Sarkar}
\affil{Department of Physics, 1 Keble Road, Oxford OX1~3NP, UK}
\beginabstract
Recent observational and theoretical developments concerning the
primordial synthesis of the light elements are reviewed, and the
implications for dark matter mentioned.
\endabstract

\section{Introduction\label{sec:intro}}

Why yet another review of big bang nucleosynthesis (BBN)? Given that
the basic physics of the synthesis of the light elements D, $^3$He,
$^4$He and $^7$Li in the big bang was thoroughly discussed over 25
years ago \cite{wfh67,w73} it is natural to wonder why so many papers
on the subject continue to appear on a regular basis. The reason for
this is two-fold. First, as observational determinations of light
element abundances have advanced, the situation has become {\em more}
instead of less uncertain. In particular it has become apparent that
inferring the primordial values of the abundances from contemporary
observations is fraught with uncertainty, given our limited
understanding of the chemical evolution of galaxies. Fortunately
observers have risen to the challenge and developed sophisticated
techniques to look further back into our past, at nearly pristine
primordial material. However large discrepancies, most likely of a
systematic nature, have subsequently emerged between different
determinations, so the improvement in precision has not led to an
increase in accuracy! Nevertheless this is a healthy development in
that it has provided a refreshing perspective on the strong claims
made in the past decade concerning the consistency of BBN predictions
with the inferred primordial abundances, and on the stringent
constraints thus inferred on new physics. The second development
concerns recent efforts to improve the accuracy of the theoretical
predictions of the abundances, and, just as important, to quantify the
uncertainties. In this talk I will discuss these issues and summarize
their implications for the dark matter problem and for new physics,
viz. $N_\nu\neq3$. Other recent reviews \cite{st98,s98,O99} may be
consulted for a different perspective than mine \cite{s96,s97}.

\section{Theoretical Calculations\label{sec:th}}

We now have excellent semi-analytic insights into the physics of BBN
which enable the $^4$He abundance to be calculated correctly to within
$\sim1-2\%$ \cite{bbf89} and the `left-over' abundances of D, $^3$He,
and $^7$Li to be estimated to within a factor of $\sim2-3$
\cite{esd91}. However for precision work, and in particular, to
determine the uncertainties, it is necessary to use the Wagoner
computer code \cite{w73} which has been improved and updated with the
latest values of the nuclear cross-sections and made freely available
to the community \cite{k92,mysite}.

There have been two important developments on this front. First the
rates for the weak interactions which determine the neutron-to-proton
ratio have been carefully calculated beyond the Born approximation,
with allowance for all zero and finite temperature radiative, Coulomb
and finite (nucleon) mass corrections \cite{lt98,emmp99}. With further
allowance for finite temperature QED corrections to the equation of
state of the plasma and the residual coupling of the neutrinos to the
plasma during $e^+e^-$ annihilation, the $^4$He abundance has been
computed with an accuracy of $\sim0.4\%$ \cite{lt98}, where the
dominant source of uncertainty comes from the present experimental
determination of the neutron lifetime: $\tau_n=885.3\pm2.0$~s
\cite{pdg98}. This is an impressive feat and it is believed that all
relevant physical processes have now been consistently taken into
account.

The uncertainties in the abundances of the other light elements are of
course considerably greater and are, moreover, correlated with each
other. The standard practice has been to use the Monte Carlo method to
sample the error distributions of the relevant reaction cross-sections
which are then input into the numerical code, thus enabling
well-defined confidence levels to be attached to the theoretically
predicted abundances \cite{skm93,kernankrauss}. However this is
computationally expensive and, moreover, needs to be repeated each
time any of the input parameters are changed or updated.\footnote{For
example, it has recently been argued \cite{bntt99} that uncertainties
in the cross-sections for some key nuclear reactions are in fact
smaller than were estimated earlier \cite{skm93}.} To overcome this
handicap we have developed a method based on linear error propagation
which requires the numerical code to be run just once to determine the
covariance (or error) matrix; simple $\chi^2$ statistics can then be
used (rather than maximum likelihood methods as with Monte Carlo
simulations) to determine e.g. the best-fit value of
$\eta\equiv\,n_{\rm N}/n_\gamma$, the nucleon-to-photon ratio
\cite{flsv98}. This method has recently been extended to consider
departures from the standard model, viz. an effective number of
neutrinos $N_\nu$ during BBN different from 3, so that correlated
limits on $\eta$ and $N_\nu$ can be extracted for any given set of
input abundances \cite{lsv99}. The results agree well, where
comparison is possible, with similar exercises using the Monte Carlo
plus maximum likelihood method \cite{ot97,hkkm98}. All the
calculations have been encoded in a simple code which is available
from a website \cite{mysite}.

The method is briefly as follows. First define the four relevant
elemental abundances as $Y_2\equiv\,$D/H, $Y_3\equiv\,^{3}$He/H,
$Y_7\equiv\,^{7}$Li/H, and finally $Y_4$ as the {\em mass fraction} of
$^4$He (often termed $Y_{\rm P}$). These depend both on the model
parameters $\eta$ and $N_\nu$, and on a network of nuclear reactions
$R_k$:
\begin{equation}
 Y_i = Y_i (\eta,\,N_\nu,\,\ldots;\,\lbrace R_k\rbrace)\ .
\end{equation}
The rates for the 12 essential nuclear reactions which have to be
considered plus their associated uncertainties have been discussed in
detail \cite{skm93} and we use these, apart from the updated value of
the neutron lifetime \cite{pdg98}. Now for a small change of the input
rate $R_k$ $(R_k\to R_k+\delta\,R_k)$, the corresponding deviation of
the $i$-th elemental abundance $(Y_i\to Y_i+\delta\,Y_i)$ as given by
linear propagation reads
\begin{equation}
 \delta\,Y_i (\eta) = Y_i(\eta) \sum_k 
  \lambda_{ik}(\eta)\frac{\delta\,R_k}{R_k}\ ,
\label{dYi}
\end{equation}
where the functions $\lambda_{ik}(\eta)$ represent the logarithmic
derivatives of $Y_i$ with respect to $R_k$:
\begin{equation}
 \lambda_{ik}(\eta) = 
  \frac{\partial\ln Y_i(\eta)}{\partial\ln R_k(\eta)}\ .
\label{lambdaik}
\end{equation}
In general, the deviations $\delta\,Y_i$ are correlated, since they
all originate from the same set of reaction rate shifts
$\{\delta\,R_k\}$. The global information is contained in the error
matrix (also called covariance matrix) \cite{edjrs71}, which is a
generalization of the ``error vector'' $\delta\,Y_i$ in
eq.~(\ref{dYi}). In particular, the abundance error matrix
$\sigma^2_{ij}(\eta)$ obtained by linearly propagating the input
$\pm1\sigma$ reaction rate uncertainties $\pm\Delta\,R_k$ to the
output abundances $Y_i$ reads:
\begin{equation}	
 \sigma^2_{ij}(\eta) = Y_i(\eta) Y_j(\eta) \sum_k
  \lambda_{ik}(\eta)\lambda_{jk}(\eta) 
  \left(\frac{\Delta\,R_k}{R_k}\right)^2\ .
\label{sigmaij}
\end{equation}
This matrix completely defines the abundance uncertainties. In
particular, the $1\sigma$ abundance errors $\sigma_i$ of $Y_i$ are
given by the square roots of the diagonal elements,
\begin{equation}
 \sigma_i(\eta)=\sqrt{\sigma^2_{ii}(\eta)}\ ,
\label{sigmaii}
\end{equation}
while the error correlations  $\rho_{ij}$ can be derived from 
eqs.~(\ref{sigmaij},\ref{sigmaii}) through the standard definition
\begin{equation}
 \rho_{ij}(\eta) = 
  \frac{\sigma_{ij}^2(\eta)}{\sigma_i(\eta)\,\sigma_j(\eta)}\ .
\label{rhoij}
\end{equation}
We have checked that the propagation of errors is indeed linear and
then computed the error matrix above, finding good agreement with the
results obtained by Monte Carlo \cite{skm93,kernankrauss}.

Next we must input the inferred values of the primordial abundances,
$\overline{Y_i}\pm\overline\sigma_i$. Taking the errors
$\overline\sigma_i$ in the determinations of different abundances
$\overline Y_i$ to be uncorrelated, the experimental squared error
matrix $\overline\sigma^2_{ij}$ is simply
\begin{equation}
 \overline \sigma^2_{ij} = \delta_{ij}\overline\sigma_i\overline\sigma_j\ ,
\label{sigmaijexp}
\end{equation}
where $\delta_{ij}$ is Kronecker's delta. The total (experimental +
theoretical) error matrix $S^2_{ij}$ is then obtained by summing the
matrices in eqs.~(\ref{sigmaij},\ref{sigmaijexp}):
\begin{equation}
 S^2_{ij}(\eta) = \sigma^2_{ij}(\eta) + \overline\sigma^2_{ij}\ .
\end{equation}
Its inverse defines the weight matrix $W_{ij}(\eta)$:
\begin{equation}
 W_{ij}(\eta) = [S^2_{ij}(\eta)]^{-1} .
\label{W}
\end{equation}
The $\chi^2$ statistic associated with the difference between
theoretical $(Y_i)$ and observational $(\overline{Y_i})$ light element
abundance determinations is then \cite{edjrs71}:
\begin{equation}
 \chi^2(\eta) = \sum_{ij} [Y_i(\eta)-\overline{Y_i}]\cdot W_{ij}(\eta)\cdot
  [Y_j(\eta)-\overline{Y_j}]\ .
\label{chi}
\end{equation}
Contours of equal $\chi^2$ can then be used to set bounds on the
parameters $\eta$ and $N_\nu$ at selected confidence levels. For
standard BBN we set $N_\nu=3$ and minimization of the $\chi^2$ then
gives the most probable value of $\eta$, while the intervals defined
by $\chi^2=\chi^2_{\rm min}+\Delta\,\chi^2$ give the likely ranges of
$\eta$ at the confidence level set by $\Delta\,\chi^2$.

Below we have provided a table of polynomial fits to the central
values of the abundances, $Y_i=a_0 + a_1 x + a_2 x^2 + a_3 x^3 + a_4
x^4 +a_5 x^5$, with $x\equiv\log_{10}(\eta/10^{-10})$ in the range
0--1 \cite{flsv98}. The value of $Y_4$ from the Wagoner code was
corrected as described earlier \cite{s96} and is in excellent
agreement with a subsequent precision calculation \cite{lt98}. These
fits, along with similar fits to the logarithmic derivatives
$\lambda_{ik}(x)$ \cite{flsv98}, have been encoded in a javascript
program which plots the abundances with associated uncertainties and
allows the corresponding values of $\eta$ to be read off for a given
input abundance \cite{seattlesite}.  We now proceed to discuss the
observational situation.

\begin{table}[b]
\begin{center}
\footnotesize\rm
\begin{tabular}{ccccccc}
\topline
    	       &  $a_0$  &  $a_1$  &  $a_2$  &  $a_3$  &  $a_4$  &  $a_5$  \\
\midline
$Y_2\times10^3$&$+0.4808$&$-1.8112$&$+3.2564$&$-3.3525$&$+1.8834$&$-0.4458$\\
$Y_3\times10^5$&$+3.4308$&$-6.1701$&$+8.1311$&$-9.7612$&$+7.7018$&$-2.5244$\\
$Y_4\times10^1$&$+2.2305$&$+0.5479$&$-0.6050$&$+0.6261$&$-0.3713$&$+0.0949$\\
$Y_7\times10^9$&$+0.5369$&$-2.8036$&$+7.6983$&$-12.571$&$+12.085$&$-3.8632$\\
\bottomline
\end{tabular}
\end{center}
\end{table}

\newpage
\section{Observed Abundances\label{sec:obs}}

The abundances of the light elements synthesized in the big bang have
been subsequently modified through chemical evolution of the
astrophysical environments where they are measured \cite{p97}. The
observational strategy then is to identify sites which have undergone
as little chemical processing as possible and rely on empirical
methods to infer the primordial abundance. For example, measurements
of deuterium (D) can now be made in quasar absorption line systems
(QAS) at high red shift; if there is a ``ceiling'' to the abundance in
different QAS then it can be assumed to be the primordial value, since
it is believed that there are no astrophysical sources of D. The
helium ($^4$He) abundance is measured in H~II regions in blue compact
galaxies (BCGs) which have undergone very little star formation; its
primordial value is inferred either by using the associated nitrogen
or oxygen abundance to track the stellar production of helium, or by
simply averaging over the most metal-poor objects \cite{h98}. (We do
not consider $^3$He which can undergo both creation and destruction in
stars \cite{p97} and is thus unreliable for use as a cosmological
probe.) Closer to home, the observed uniform abundance of lithium
($^7{\rm Li}$) in the hottest and most metal-poor Pop~II stars in our
Galaxy is believed to reflect its primordial value \cite{m97}.

However as remarked earlier, improvements in observational methods
have made the situation more, instead of less, uncertain. Large
discrepancies, of a systematic nature, have emerged between different
observers who report, e.g., relatively `high' \cite{dhi,rh96,dhinew}
or `low' \cite{dlo} values of deuterium in different QAS, and `low'
\cite{he4old,oss97} or `high' \cite{he4new,it98} values of helium in
BCG, using different data reduction methods. It has been argued that
the Pop~II lithium abundance \cite{t94,mpb95,bm97} may in fact have
been significantly depleted down from its primordial value
\cite{pdd92,vc9598}, with some observers arguing to the contrary
\cite{bm98}.

Rather than take sides in this matter we instead consider several
combinations of observational determinations, which cover a wide range
of possibilities, in order to demonstrate our method and obtain
illustrative best-fits for $\eta$ and $N_\nu$. 

\subsection{Deuterium}
 
The first observations of D absorption due to a QAS at redshift
$z=3.32$ towards Q0014+813 suggested a relatively `high' value of
\cite{dhi,rh96}
\begin{equation}
 \overline{Y_2} = 1.9 \pm 0.4 \times 10^{-4}\ ,
\label{dhi}
\end{equation}
and was consistent with limits set in other QAS. However subsequent
independent measurements in QAS at $z=3.572$ towards Q1937-1009 and at
$z=2.504$ towards Q1009+2956 found a much lower abundance of
\cite{dlo}
\begin{equation}
 \overline{Y_2} = 3.4 \pm 0.3 \times 10^{-5}\ .
\label{dlo}
\end{equation}
More recently, observations of a QAS at $z=0.701$ towards Q1718+4807
have also yielded a high abundance of \cite{dhinew}
\begin{equation}
 \overline{Y_2} = 3.3 \pm 1.2 \times 10^{-4}\ .
\label{dhinew}
\end{equation}

These discrepant measurements may result from spatial variations of
the D abundance due to localized astrophysical processes \cite{rh96}
or due to inhomogeneous nucleosynthesis \cite{jf96}. More prosaic
explanations invoke systematic effects. For example, new data on the
Q0014+813 system \cite{tbk96} reveal the presence of an `interloper'
hydrogen line contaminating the deuterium absorption feature. It has
also been suggested \cite{mesoturb} that the discordance between the
`high' and `low' values may be considerably reduced if the analysis of
the H+D profiles accounts for the correlated velocity field of bulk
motion, i.e. mesoturbulence, rather than being based on
multi-component microturbulent models. It is then found that a true
primordial abundance of
\begin{equation}
 \overline{Y_2} = (3.5 - 5.2) \times 10^{-5} 
\label{dcomp}
\end{equation}
is compatible simultaneously (at 95\% C.L.) with the `low' and `high'
values quoted above in eqs.~(\ref{dlo},\ref{dhinew}) if the data on
these objects is appropriately analysed. Whether this is indeed the
correct explanation in all cases is however not clear.  Obviously more
measurements are needed and at the moment all the suggested values of
the primordial deuterium abundance quoted above merit consideration.

\subsection{Helium}

The primordial helium abundance is found to be \cite{oss97}
\begin{equation}
 \overline{Y_4} = 0.234 \pm 0.002 \pm 0.005 \ ,
\label{he4old}
\end{equation}
from linear regression to zero metallicity in a set of 62 BCGs, based
largely on observations which gave a relatively `low' value
\cite{he4old}. However using data on a sample of 27 BCGs from an
idependent survey, a significantly higher value of
\begin{equation}
 \overline{Y_4} = 0.243 \pm 0.003 \ ,
\label{he4new}
\end{equation} 
has been derived from a new analysis which uses the helium lines
themselves to self-consistently determine the physical conditions in
the H~II region \cite{he4new} and specifically excludes those regions
(with anomalously low He I line intensities) which are believed to be
affected by underlying stellar absorption. In particular it has been
demonstrated that there is strong underlying stellar absorption in the
NW component of I~Zw~18, which was included in earlier analyses
\cite{he4old}.  Excluding this, the average of the values found in the
two most metal-poor BCGs, I~Zw~18 and SBS~0335-052, is \cite{it98}
\begin{equation}
 \overline{Y_4} = 0.245 \pm 0.004 \ .
\label{he4newer}
\end{equation}
Again further work is needed to resolve the situation but we will
consider both the `low' and `high' values for the primordial helium
abundance.

\subsection{Lithium}

The controversy here concerns whether the observed approximately
constant `Spite plateau' in the abundance of $^7$Li in Pop~II stars in
the halo reflects its true primordial value or has been depleted down
from an initially higher abundance. There is agreement that the average
value in the hottest and most metal-poor stars is about
\cite{t94,mpb95}
\begin{equation}
 \overline{Y_7} = 1.6 \pm 0.36 \times 10^{-10} \ ,
\label{li7}
\end{equation}
but there is dispute about whether there is significant scatter
suggestive of some depletion mechanism, or a slight trend of
decreasing $^7$Li abundance with increasing metallicity suggestive of
secondary cosmic ray production. A recent analysis based on 41 stars
finds no evidence of any depletion or post-BBN creation and suggests a
slightly higher primordial abundance \cite{bm97},
\begin{equation}
 \overline{Y_7} = 1.73 \pm 0.21 \times 10^{-10} \ .
\label{li7new}
\end{equation}
(The systematic error may have to be raised by $\sim50\%$ to allow for
the uncertainty in the oscillator strengths of the lithium lines
\cite{m97}.)

However there are hot, metal-poor Pop~II stars which appear identical
to those which define the `Spite plateau' but which have {\em no}
detectable $^7$Li \cite{t94,rnb98}, indicating that some depletion
mechanism does exist. Recently it has been argued that the observed
abundance has been depleted down from a primordial value of
\cite{pwsn98}
\begin{equation}
 \overline{Y_7} = 3.9 \pm 0.85 \times 10^{-10} \ ,
\label{li7depl}
\end{equation}
the lower end of the range being set by the presence of the highly
overdepleted halo stars and consistency with the $^7$Li abundance in
the Sun and in open clusters, while the upper end of the range is set
by the observed dispersion about the `Spite plateau' and the
$^6$Li/$^7$Li depletion ratio. A somewhat smaller depletion is
suggested by other workers \cite{vc9598} who find a primordial
abundance of $\overline{Y_7}=2.3\pm0.5\times10^{-10}$.

The implications of all these possibilities need study.

\section{Implications for $\eta$ and baryonic dark matter \label{sec:eta}}

Elsewhere \cite{lsv99} we have discussed various combinations of the
above data sets and provided a handy computer programme \cite{mysite}
to enable the reader to consider other permutations/combinations.
Here we will present results for just two possible (although mutually
incompatible) selections of measurements which we name data set ``A''
and data set ``B'':
\begin{equation}
{\rm Data\  set\  A:\ }\left\{
\begin{array}{l}
 \overline{Y_2} = 1.9 \pm 0.4 \times 10^{-4}  \ , \\
 \overline{Y_4} =  0.234 \pm 0.0054             \ , \\
 \overline{Y_7} =  1.6 \pm 0.36 \times 10^{-10}\ ; 
\end{array}\right.
\label{datasetA}
\end{equation}
\begin{equation}
{\rm Data\  set\  B:\ }\left\{
\begin{array}{l}
 \overline{Y_2} =  3.40 \pm 0.25 \times 10^{-5}  \ , \\
 \overline{Y_4} =  0.243 \pm 0.003             \ , \\
 \overline{Y_7} =  1.73 \pm 0.21 \times 10^{-10}\ . 
\end{array}\right.
\label{datasetB}
\end{equation}
(In subsequent figures we also consider a slight variation of data set
B in which we use eq.~(\ref{he4newer}) instead of eq.~(\ref{he4new})
for the $^4$He abundance.)

\begin{figure}[h!tb]
\epsfxsize\hsize
\epsfbox{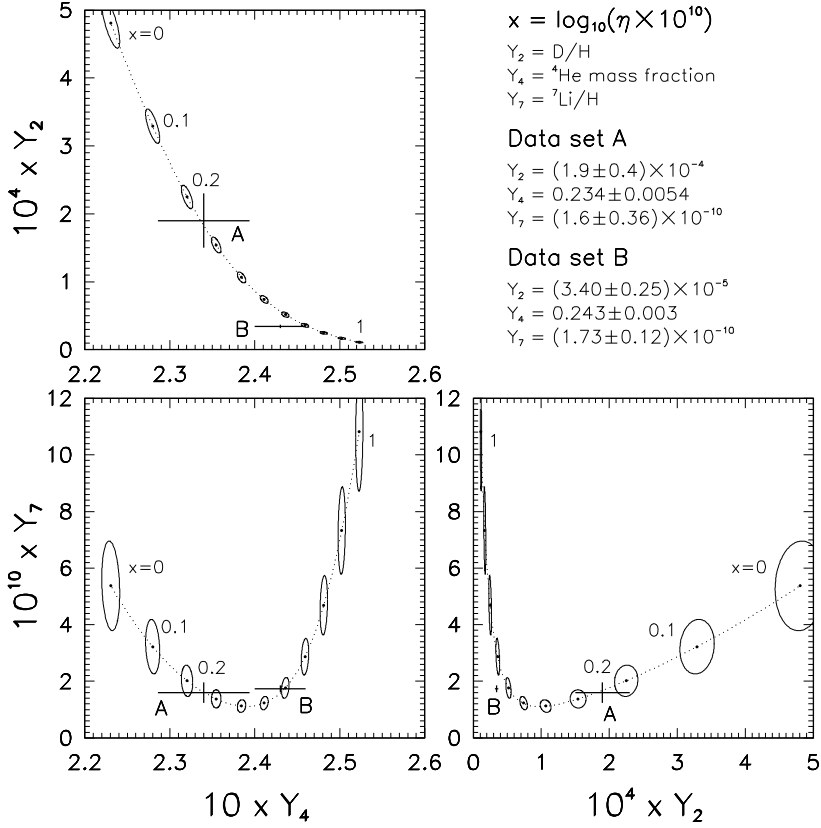}
\caption{Standard BBN predictions (dotted lines) in the planes defined
 by the abundances $Y_2$, $Y_4$, and $Y_7$ \protect\cite{flsv98}. The
 theoretical uncertainties are depicted as $1\sigma$ error ellipses at
 $x\equiv\log_{10}(\eta/10^{-10}=0,\,0.1,\,0.2,\,\ldots,\,1$. The
 crosses indicate the two observational data sets (with $1\sigma$
 errors).}
\label{crosscorr}
\end{figure}

Figure~\ref{crosscorr} shows the theoretical predictions compared with
the data. We calculate and minimize the $\chi^2$ and obtain the 95\%
C.L. ranges allowed by each data set by cutting the curves at
$\Delta\chi^2=\chi^2-\chi^2_{\rm min}= 3.84$:
\begin{eqnarray}
 {\rm Data\ set\ A:\ }\ \eta & = & 1.78^{+0.54}_{-0.34} \times 10^{-10} ,\\
 {\rm Data\ set\ B:\ }\ \eta & = & 5.13^{+0.72}_{-0.66} \times 10^{-10} .
\end{eqnarray}

The range of $\eta$ for data set A agrees very well with the 95\%
C.L.\ range obtained independently with the same input data but using
the Monte Carlo plus maximum likelihood method \cite{ot97}. However it
is inconsistent with a further constraint on $\eta$ coming from a
recent analysis of the Ly$\alpha$-``forest'' absorption lines in
quasar spectra. The observed mean opacity of the lines requires some
minimum amount of neutral hydrogen in the high redshift intergalactic
medium, given a lower bound to the flux of ionizing radiation. Taking
the latter from a conservative estimate of the integrated UV
background due to quasars yields the constraint \cite{lyalpha}
\begin{equation}
 \eta > 3.4 \times 10^{-10}\ . 
\label{lya}
\end{equation}
Using $\Omega_{\rm N}h^2\simeq\eta/2.73\times10^{-8}$ this corresponds
to $\Omega_{\rm N}>0.0125h^{-2}$, where $\Omega_{\rm N}$ is the
density of nucleons in ratio to the critical density and $h$ is the
present Hubble parameter in units of $100$~km~s$^{-1}$~Mpc$^{-1}$.

Thus data set B is favoured by the Ly$\alpha$-forest
constraint. The implied nucleon density is significantly higher than
the nucleonic matter in stars and X-ray emitting gas in groups and
clusters today \cite{baryon},
\begin{equation}
 \Omega_{\rm N} \equiv \frac{\rho_{\rm N}}{\rho_{\rm c}} \sim 0.0025
 h^{-1} + 0.0046 h^{-1.5} \ ,
\end{equation}
implying that most of the nucleons are presently in some dark form.

\newpage
\section{New Physics and the Possibility of $N_\nu\neq3$ \label{sec:nnu}}

We should also take into account that the goodness of fit can be
affected by a non-standard Hubble expansion rate during BBN, e.g. due
to the presence of new neutrinos. Although the Standard Model contains
only $N_\nu=3$ weakly interacting massless neutrinos, the recent
experimental evidence for neutrino oscillations \cite{nuoscrev} may
require it to be extended to include new superweakly interacting
massless (or very light) particles such as singlet neutrinos or
Majorons. These do not couple to the $Z^0$ vector boson and are
therefore unconstrained by the precision studies of $Z^0$ decays which
establish the number of $SU(2)_{\rm L}$ doublet neutrino species to be
\cite{pdg98}
\begin{equation}
 N_{\nu} = 2.993 \pm 0.011 \ .
\label{nnulep}
\end{equation}
However, as was emphasized some time ago \cite{history}, such
particles would boost the relativistic energy density, hence the
expansion rate, during BBN, thus increasing the yield of $^4$He. This
argument was quantified for new types of neutrinos and new superweakly
interacting particles \cite{chicago} in terms of a bound on the {\em
equivalent number of massless neutrinos} present during
nucleosynthesis:
\begin{equation}
 N_{\nu} = 3 + f_{\rm B, F} \sum_{i} \frac{g_{i}}{2} 
                                     \left(\frac{T_{i}}{T_{\nu}}\right)^4\ ,
\label{Nnudef}
\end{equation}
where $g_i$ is the number of (interacting) helicity states, $f_{\rm
B}=8/7$ (bosons) and $f_{\rm F}=1$ (fermions), and the ratio
$T_{i}/T_{\nu}$ depends on the thermal history of the particle under
consideration \cite{oss81}. For example, $T_{i}/T_{\nu}\leq0.465$ for
a particle which decouples above the electroweak scale such as a
singlet Majoron or a sterile neutrino. However the situation may be
more complicated, e.g. if the sterile neutrino has large mixing with a
left-handed doublet species, it can be brought into equilibrium
through (matter-enhanced) oscillations in the early universe, making
$T_{i}/T_{\nu}\simeq1$ \cite{ekt92}. Moreover such oscillations can
generate an asymmetry between $\nu_{\rm e}$ and $\bar{\nu_{\rm e}}$,
thus directly affecting neutron-proton interconversions and the
resultant yield of $^4$He \cite{nuosc}. This can be quantified in
terms of the {\em effective} value of $N_\nu$ parametrizing the
expansion rate during BBN, which may well be below 3! Similarly,
non-trivial changes in $N_\nu$ can be induced by the decays
\cite{nudec} or annihilations \cite{nuann} of massive neutrinos (into
e.g. Majorons), so it is clear that it is a sensitive probe of new
physics.

The values of $N_\nu$ and $\eta$ are correlated since the effect of a
faster expansion rate can be compensated for by the effect of a
smaller nucleon density. Indeed increasingly stringent lower bounds on
$\eta$ inferred from assumed upper limits to the primordial deuterium
abundance (which were deduced from chemical evolution arguments) have
been used to set increasingly stringent upper bounds on $N_\nu$. These
have ranged from 4 downwards \cite{many}, culminating in one below 3
which precipitated the so-called ``crisis'' for standard BBN
\cite{crisis}, and was interpreted as requiring new physics. However
as cautioned before \cite{eens86}, there are large systematic
uncertainties in such constraints on $N_\nu$ which are sensitive to
our limited understanding of galactic chemical evolution. Moreover it
has been emphasized \cite{kernankrauss} that the procedure used
earlier \cite{many} to bound $N_\nu$ was statistically inconsistent
since, e.g., correlations between the different elemental abundances
were not taken into account. This can of course be addressed by the
Monte Carlo method, using which it was shown \cite{nocrisis} that the
{\em conservative} observational limits on the primordial abundances
of D, $^4$He and $^7$Li allow $N_\nu\leq4.53$ ($95\%$ C.L.),
significantly less restrictive than earlier estimates. However because
the Monte Carlo method is computationally expensive and the exercise
needs to be redone e.g. whenever the input abundances change, it would
clearly be advantageous to extend our linear error propagation
approach to include $N_\nu$ as a parameter.

This is, in principle, straightforward, since it simply requires
recalculation of the functions $Y_i$ and $\lambda_{ik}$ at the chosen
value of $N_\nu$. Since it would be impractical to use extensive
tables of polynomial coefficients,we have devised some formulae which,
to good accuracy, relate the calculations for arbitrary values of
$N_\nu$ to the standard case $N_\nu=3$, thus reducing the numerical
task dramatically \cite{lsv99}. 

As is known from previous work \cite{esd91}, the synthesized elemental
abundances ${\rm D}/{\rm H}$, $^3{\rm He}/{\rm H}$, and $^7{\rm
Li}/{\rm H}$ (i.e., $Y_2$, $Y_3$, and $Y_7$ in our notation) are given
to a good approximation by the quasi-fixed points of the corresponding
rate equations, which formally read
\begin{equation}
 \frac{{\rm d}Y_i}{{\rm d}t} \propto \eta\, \sum_{+,-} Y \times Y \times 
  \langle\sigma v\rangle_{\rm T}\ ,
\end{equation}
where the sum runs over the relevant source $(+)$ and sink $(-)$
terms, and $\langle \sigma v\rangle_T$ is the thermally-averaged
reaction cross section. Since the temperature of the universe evolves
as ${\rm d}T/{\rm d}t\propto-T^3\sqrt{g_\star}$, with the number of
relativistic degrees of freedom, $g_\star=2+(7/4)(4/11)^{4/3}N_{\nu}$
(following $e^+e^-$ annihilation), the above equation can be rewritten
as
\begin{equation}
 \frac{{\rm d}Y_i}{{\rm d}T} \propto -\frac{\eta}{g_\star^{1/2}}
 \,T^{-3}\, \sum_{+,-}  Y \times Y \times \langle\sigma v\rangle_{\rm T}\ ,
\end{equation}
which shows that $Y_2$, $Y_3$, and $Y_7$ depend on $\eta$ and $N_\nu$
essentially through the combination $\eta/g_\star^{1/2}$. Thus the
calculated abundances $Y_2$, $Y_3$, and $Y_7$ (as well as their
logarithmic derivatives $\lambda_{ik}$) should be approximately
constant for
\begin{equation}
 \log \eta -\frac{1}{2} \log g_\star = {\rm constant}\ ,
\label{shift}
\end{equation}
as we have verified numerically.

This suggests that the values of $Y_i$ and of $\lambda_{ik}$ for
$N_\nu=3+\Delta{N_\nu}$ can be related to the case $N_\nu=3$ through
an appropriate shift in $x$:
\begin{eqnarray}
 Y_i(x,3+\Delta{N_\nu}) &\simeq& Y_i(x+c_i\Delta{N_\nu},3)\ ,\\
 \lambda_{ik}(x,3+\Delta{N_\nu}) &\simeq& \lambda_{ik}(x+c_i\Delta{N_\nu},3)\ ,
\end{eqnarray}
where the coefficient $c_i$ is estimated to be $\sim-0.03$ from
eq.~(\ref{shift}) (at least for small $\Delta{N_\nu}$). In order to
obtain a satisfactory accuracy in the whole range
$(x,N_\nu)\in[0,1]\times [1,5]$, we in fact allow upto a second-order
variation in $\Delta{N_\nu}$, and for a rescaling factor of the
$Y_i$'s. As regards the $^4{\rm He}$ abundance, a semi-analytical
approximation \cite{bbf89} also suggests a relation between $x$ and
$N_\nu$ similar to eq.~(\ref{shift}), although with different
coefficients, and indeed functional relations of the kind above work
well also in this case. However, in order to achieve higher accuracy
and, in particular, to match exactly the result of the recent
precision calculation of $Y_4$ which includes all finite temperature
and finite density corrections \cite{lt98}, we also allow for a
rescaling factor for the $\lambda_{4k}$'s. All these calculations have
been embedded in a compact Fortran programme available from a website
\cite{mysite}, which allows easy extraction of joint fits to $x$ and
$N_\nu$ for a given set of elemental abundances, without having to run
the full BBN code, and with no significant loss in accuracy. We now
proceed to discuss these fits.

\section{Joint fits to $\eta$ and $N_\nu$ \label{sec: joint}}

\begin{figure}[h!tb]
\epsfxsize\hsize
\epsfbox{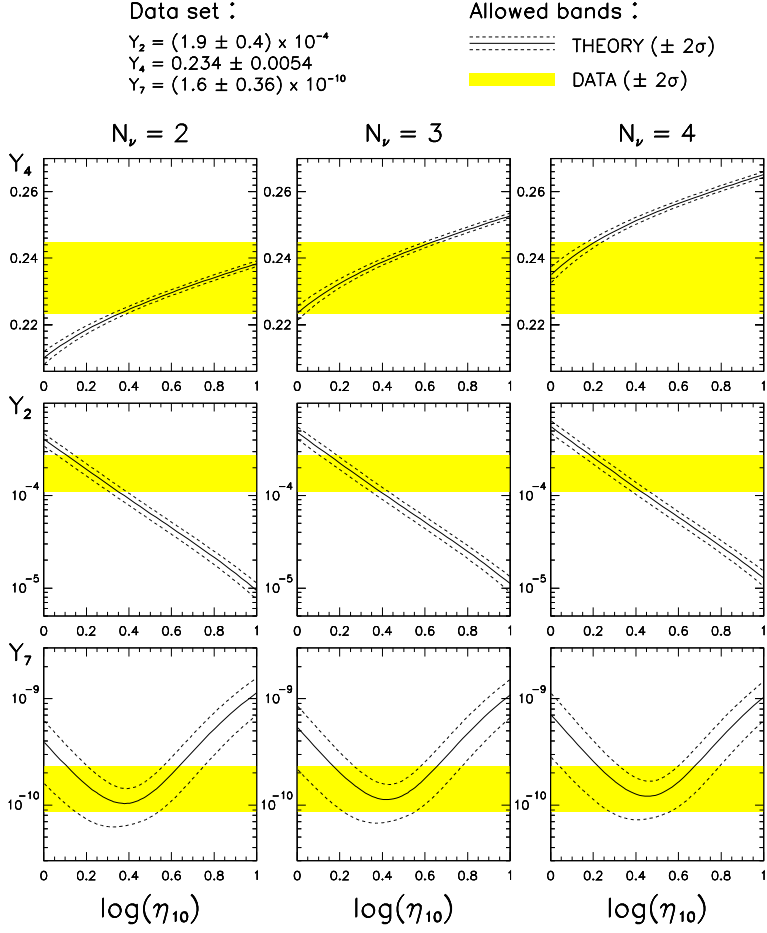}
\caption{Primordial abundances $Y_4$ ($^4$He mass fraction) $Y_2$
 (D/H) and $Y_7$ ($^7$Li/H), for $N_\nu=2$, 3, and 4. Solid and dashed
 curves represent the theoretical central values and the $\pm2\sigma$
 bands, respectively. The grey areas represent the $\pm2\sigma$
 experimental bands for the data set A \protect\cite{lsv99}.}
\label{abundA}
\end{figure}

\begin{figure}[h!tb]
\epsfxsize\hsize
\epsfbox{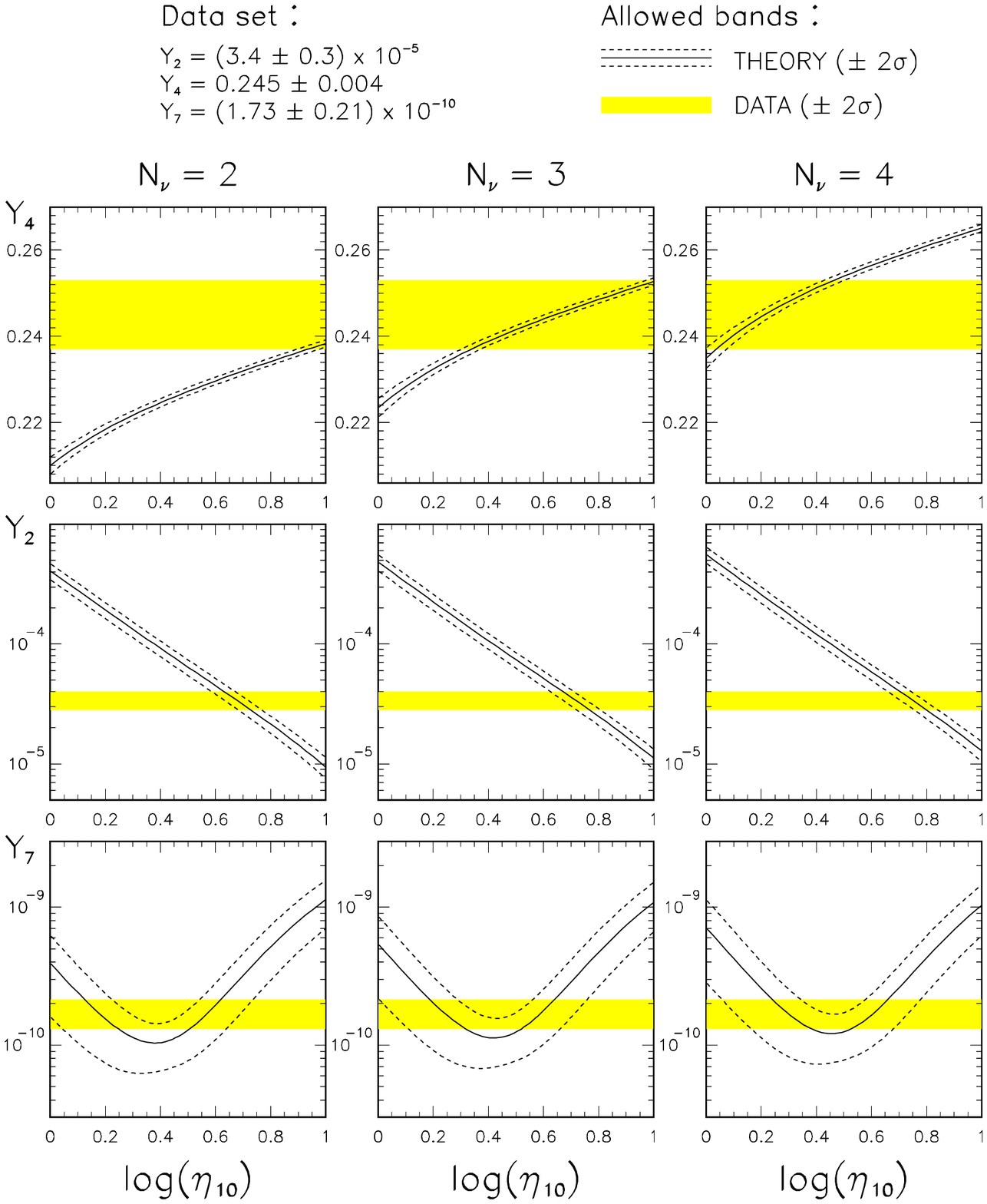}
\caption{Same as in figure~\protect\ref{abundA}, but for data set
 B \protect\cite{lsv99}.}
\label{abundB}
\end{figure}

Figure~\ref{abundA} shows the abundances for various $N_\nu$ compared
for illustration with data set A. While there is consistency between
theory and data for $x\sim0.2-0.4$ and $N_\nu=3$, for $N_\nu=2$
($N_\nu=4$) the $Y_2$ data prefer values of $x$ lower (higher) than
the $Y_4$ data. Therefore, we expect that a global fit will favor
values of $(x,N_\nu)$ close to $(0.3,3)$. Similarly, a fit to data set
B should favour values of $(x,N_\nu)\sim(0.7,3)$ as seen in
figure~\ref{abundB}.

\begin{figure}[h!tb]
\epsfxsize\hsize
\epsfbox{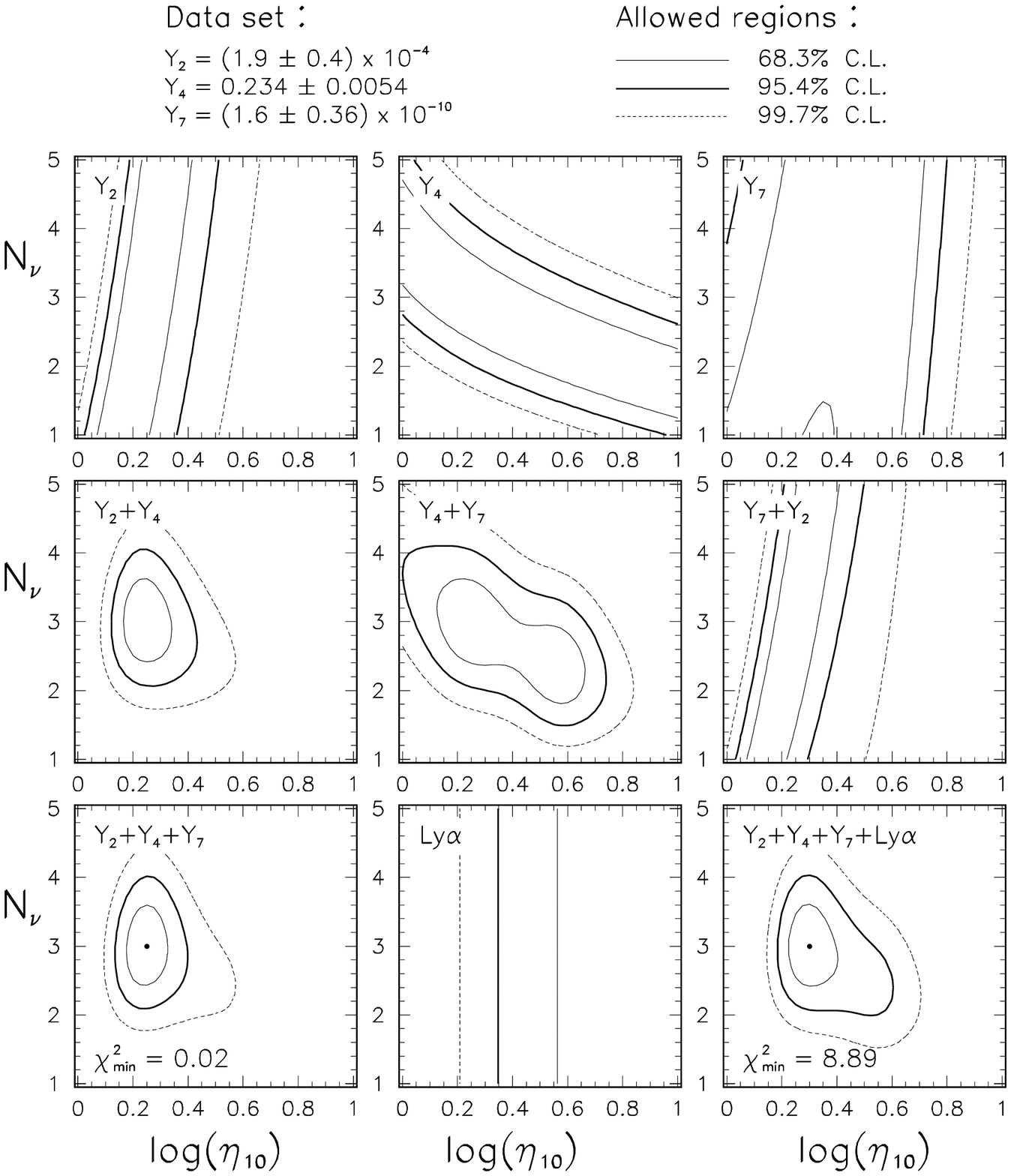}
\caption{Joint fits to $x=\log(\eta_{10})$ and $N_\nu$ using the
 abundances of data set A \protect\cite{lsv99}. The abundances
 $Y_2$, $Y_4$, and $Y_7$ are used separately (upper panels), in
 combinations of two (middle panels), and all together, without and
 with the Ly$\alpha$-forest constraint on $\eta$ (lower panels).}
\label{chi2A}
\end{figure}

To quantify this we show in figure~\ref{chi2A} the results of joint
fits using the abundances of data set A. The abundances $Y_2$, $Y_4$,
and $Y_7$ are used separately (upper panels), in combinations of two
(middle panels), and all together, without and with the
Ly$\alpha$-forest constraint\footnote{This bound is not well-defined
statistically but we can parametrize it for example through a penalty
function $\chi^2_{{\rm
Ly}\alpha}(\eta)=2.7\times(3.4\times10^{-10}/\eta)^2$ which can be
added to the $\chi^2$ derived from our fit to the elemental
abundances.} on $\eta$ (lower panels). In this way the relative weight
of each piece of data in the global fit can be understood at
glance. The three C.L. curves (solid, thick solid, and dashed) are
defined by $\chi^2-\chi^2_{\rm min} = 2.3,\, 6.2$, and $11.8$,
respectively, corresponding to 68.3\%, 95.4\%, and 99.7\% C.L. for two
degrees of freedom ($\eta$ and $N_\nu$), i.e., to the probability
intervals often designated as 1, 2, and 3 standard deviation
limits. The $\chi^2$ is minimized for each combination of $Y_i$, but
the actual value of $\chi^2_{\rm min}$ (and the best-fit point) is
shown only for the relevant global combination $Y_2+Y_4+Y_7(+{\rm
Ly}\alpha)$. For this data set, the helium and deuterium abundances
dominate the fit, as can be seen by comparing the combinations
$Y_2+Y_4$ and $Y_2+Y_4+Y_7$. The preferred values of $x$ are
relatively low, and the preferred values of $N_\nu$ range between 2
and 4. Although the fit is excellent, the low value of $x$ is in
conflict with the Ly$\alpha$-forest constraint on $\eta$, as indicated
by the increase of $\chi^2_{\min}$ from 0.02 to 8.89.

\begin{figure}[h!tb]
\epsfxsize\hsize
\epsfbox{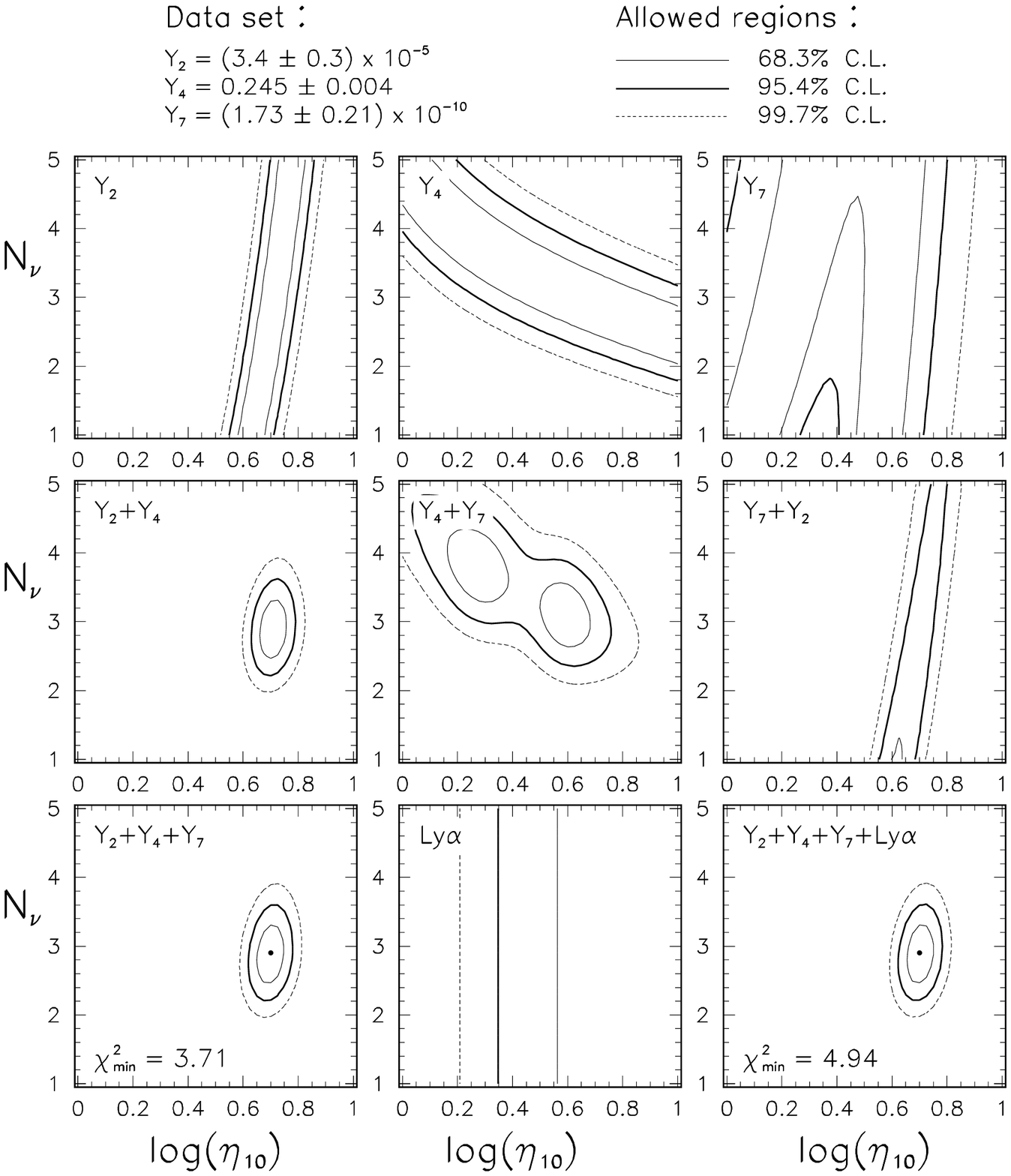}
\caption{Same as in Fig.~\ref{chi2A}, but for the data set B
 \protect\cite{lsv99}.}
\label{chi2B}
\end{figure}

As shown in figure~\ref{chi2B} there is no such problem with data set
B which favors high values of $x$ because of the `low' deuterium
abundance. The combination of $Y_2+Y_7$ isolates, at high $x$, a
narrow strip which depends mildly on $N_\nu$. The inclusion of $Y_4$
selects the central part of such strip, corresponding to $N_\nu$
between 2 and 4. As in figure~\ref{chi2A}, the combination $Y_4+Y_7$
does not appear to be very constraining. The overall fit to
$Y_2+Y_4+Y_7$ is acceptable but not particularly good, mainly because
$Y_2$ and $Y_7$ are only marginally compatible at high $x$. On the
other hand, the $Y_2+Y_4+Y_7$ bounds are quite consistent with the
Ly$\alpha$-forest constraint.

\begin{figure}[h!tb]
\epsfxsize\hsize
\epsfbox{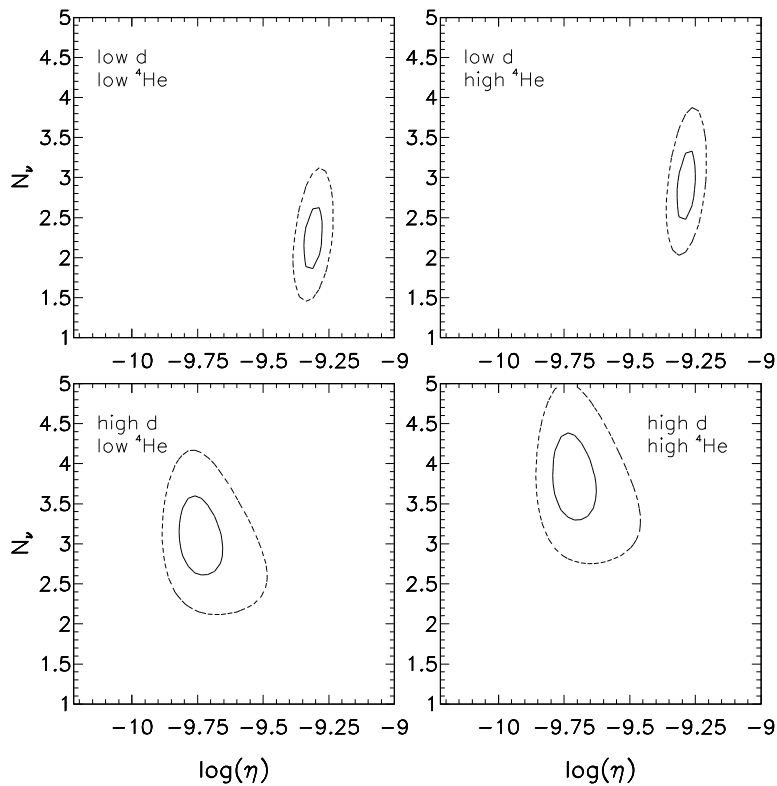}
\caption{The $68\%$ (solid) and $95\%$ (dotted) likelihood contours
 for the number of neutrino species and the nucleon-to-photon ratio,
 for all four combinations of the `high' and `low' deuterium and
 helium abundance measurements in data sets A and B.}
\label{chi2cross}
\end{figure}

Elsewhere \cite{lsv99} we have presented results for other possible
combinations of input abundances. As shown in figure~\ref{chi2cross}
(cf. \cite{hkkm98}), one can also consider orthogonal
combinations to those above, e.g. `high' deuterium {\em and} `high'
helium, or `low' deuterium {\em and} `low' helium. The latter
combination implies $N_\nu\sim2$, thus creating the so-called
``crisis'' for standard nucleosynthesis \cite{crisis}. Conversely, the
former combination suggests $N_\nu\sim4$, which would also constitute
evidence for new physics. Allowing for depletion of the primordial
lithium abundance to its Pop~II value, relaxes the upper bound on
$N_\nu$ further, as was noted earlier \cite{nocrisis}.

\section{Conclusions\label{sec:concl}}

It is clear from the above discussion that the present observational
data on the primordial elemental abundances are not as yet
sufficiently stable to derive firm bounds on $\eta$ and
$N_\nu$. Different and arguably equally acceptable choices for the
input data sets lead to very different predictions for $\eta$, and to
relatively loose constraints on $N_\nu$ in the range 2 to 4 at the
95\% C.L. Thus it may be premature to quote restrictive bounds based
on some particular combination of the observations, until the
discrepancies between different estimates are satisfactorily
resolved. Our method of analysis provides the reader with an
easy-to-use technique \cite{mysite} to recalculate the best-fit values
as the observational situation evolves further.

\begin{figure}[h!tb]
\epsfxsize\hsize
\epsfbox{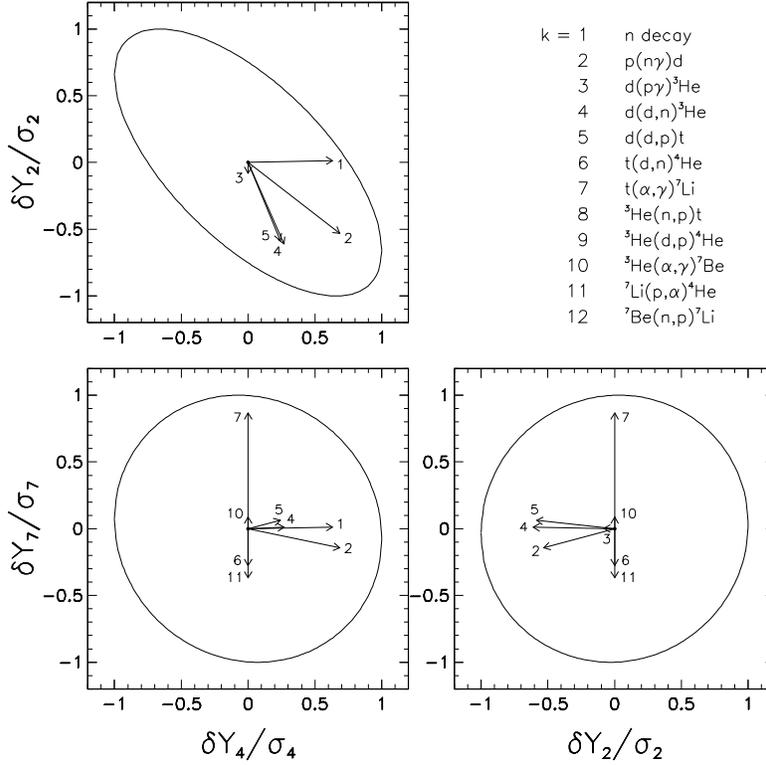}
\caption{Individual contributions of different reaction rates $R_k$ to
 the uncertainties in $Y_2$, $Y_4$, and $Y_7$, normalized to the
 corresponding total errors $\sigma_2$, $\sigma_4$, and $\sigma_7$,
 for $\eta=1.78\times10^{-10}$, the best-fit value for data set A
 \protect\cite{flsv98}. Each arrow corresponds to the shift
 $\delta\,Y_i$ induced by a $+1\sigma$ shift of $R_k$. Some small
 error components have not been plotted.}
\label{nucratA}
\end{figure}

\begin{figure}[h!tb]
\epsfxsize\hsize
\epsfbox{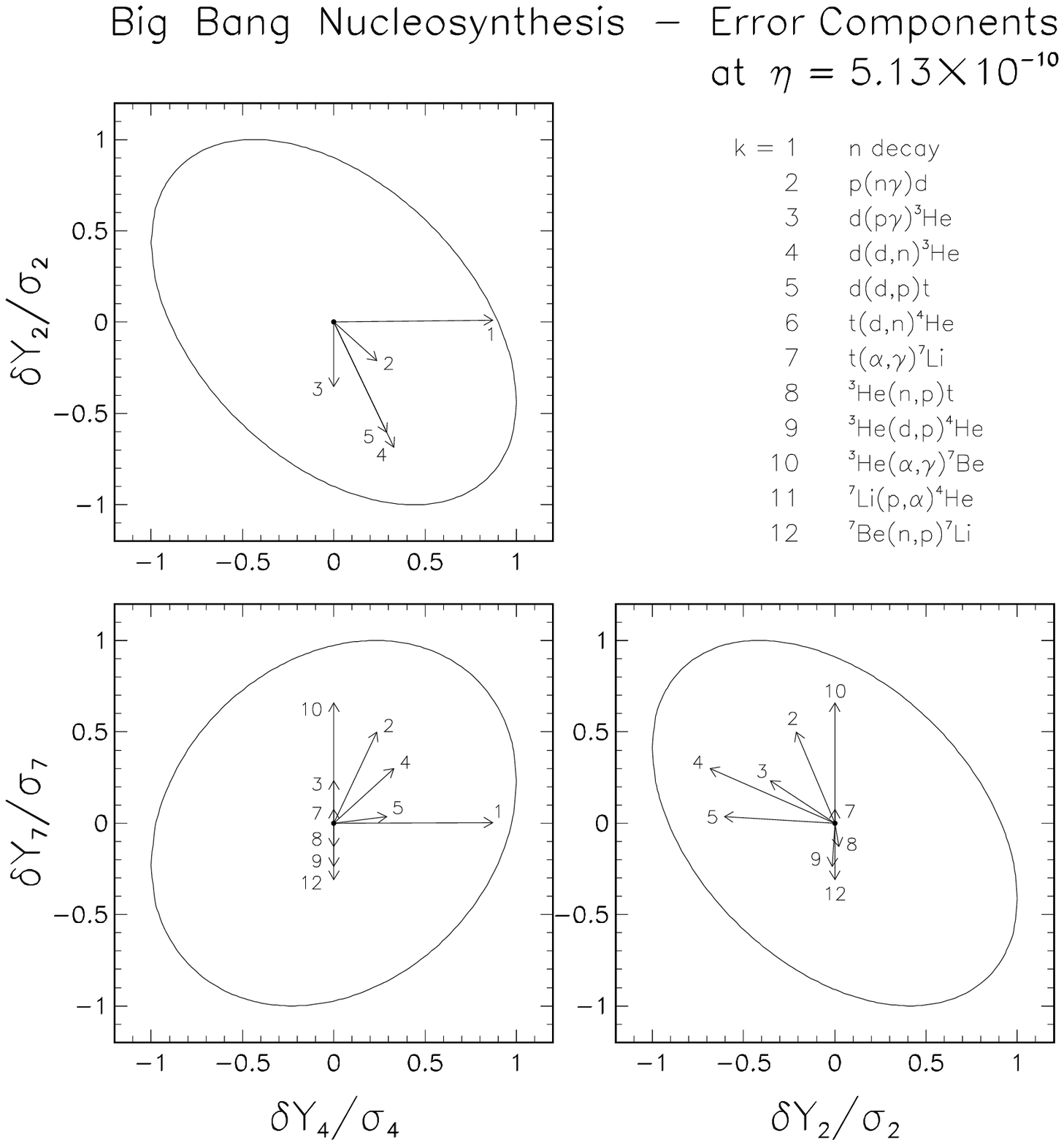}
\caption{Same as in figure~\protect\ref{nucratA}, but for
 $\eta=5.13\times10^{-10}$, the best-fit value for data set B
 \protect\cite{flsv98}.}
\label{nucratB}
\end{figure}

In order to improve the theoretical predictions further it is
necessary to know the nuclear reaction rates better. To determine
which reaction is largely responsible for the uncertainty in a
particular elemental abundance (for any given nucleon density) is
clearly important in this regard. In our approach this can be easily
determined by ``perturbing'' the values of the input reaction rates
and observing their effect on the predicted abundances, namely one can
study the contribution to the total uncertainty $\sigma_i$ of $Y_i$
induced by a $+1\sigma$ shift of $R_k$:
\begin{equation}
 R_k \to R_k+\Delta\,R_k\ \Longrightarrow\ Y_i \to Y_i + \delta\,Y_i\ .
\end{equation}
We choose two particular values of $\eta$, $1.78\times10^{-10}$ and
$5.13\times10^{-10}$, corresponding to the best-fit values for data
sets A and B respectively and show in figures~\ref{nucratA} and
\ref{nucratB} the deviations $\delta\,Y_i$ (normalized to the total
error $\sigma_i$) induced by $+1\sigma$ shifts in the $R_k$'s, plotted
in the same set of planes as used for figure~\ref{crosscorr}. The
$1\sigma$ error ellipses shown in these figures are obtained by
combining the deviation vectors $\delta Y_i/\sigma_i$ in an
uncorrelated manner. Several interesting conclusions can be drawn from
this exercise. As expected, the uncertainty in the weak interaction
rate $R_1$ has the greatest impact on $Y_4$ for the high value of
$\eta$ (figure~\ref{nucratB}), since essentially all neutrons end up
being bound in $^4$He. However at the lower value of $\eta$
(figure~\ref{nucratA}), the uncertainty in $R_2$ --- the ``deuterium
bottleneck'' --- plays an equally important role as $R_1$ in
determining $Y_4$ because nuclear burning is less complete here than
at high $\eta$. Similarly with reference to the reaction rates $R_7$,
$R_{10}-R_{12}$ which synthesize $^7$Li, at low $\eta$ it is the
competition between $R_7$ and $R_{11}$ which largely determines $Y_7$,
while at high $\eta$ it is the competition between $R_{10}$ and
$R_{12}$. The anticorrelation between $Y_4$ and $Y_2$ is driven mainly
by $R_2$ at low $\eta$ and, to a lesser extent, by $R_4$ and $R_5$,
while the reverse is the case at high $\eta$. The anticorrelation
between $Y_4$ and $Y_7$ at low $\eta$ is also basically driven by
$R_2$, while the correlation at high $\eta$ is due to both $R_2$ and
$R_4$. Thus we have a direct visual basis for assessing in what
direction the output abundances $Y_i$ are pulled by possible changes
in the input cross sections $R_k$. This should be helpful in
determining experimental strategies to determine key reaction rates
more precisely.

However one might ask what would happen if these discrepancies remain?
We have already noted the importance of an independent constraint on
$\eta$ (from the Ly$\alpha$-forest) in discriminating between
different options. However, given the many assumptions which go into
the argument \cite{lyalpha}, this constraint is rather uncertain at
present. Fortunately it should be possible in the near future to
independently determine $\eta$ to within $\sim5\%$ through
measurements of the angular anisotropy of the cosmic microwave
background (CMB) on small angular scales \cite{cmb}, in particular
with data from the all-sky surveyors MAP and PLANCK
\cite{missions}. Such observations will also provide a precision
measure of the relativistic particle content of the primordial
plasma. Hopefully the primordial abundance of $^4$He would have
stabilized by then, thus providing, in conjunction with the above
measurements. a reliable probe of a wide variety of new physics and
astrophysics which can affect nucleosynthesis (for recent work see
\cite{bfv98,dp98,ggr98,kks98,mt98,rj98}).

\section*{Acknowledgments}

It is a pleasure to thank Laura Baudis and Hans Klapdor-Kleingrothaus
for organizing this enjoyable meeting, and my collaborators on BBN,
Eligio Lisi and Francesco Villante, for allowing me to present our
unpublished results.

\end{document}